\documentclass[preprint,nofootinbib]{revtex4}%
\usepackage{tikz}
\definecolor{bluegray}{RGB}{31,95,122}
\usepackage{amssymb}
\usepackage{subcaption}
\usepackage{amsfonts}
\usepackage{amsmath}
\usepackage{amsmath}
\newcommand{\ket}[1]{\lvert #1\rangle}
\usepackage{graphicx}
\usepackage{float}
\usepackage{mathrsfs}
\providecommand{\dd}{\mathrm{d}}

\providecommand{\Vb}{V_{\mathrm B}}
\providecommand{\U}[1]{\protect\rule{.1in}{.1in}}
\providecommand{\U}[1]{\protect\rule{.1in}{.1in}}
\definecolor{blue}{rgb}{0,0,1}

\definecolor{red}{rgb}{1,0,0}

\begin{document}
\title{Holographic probes and Krylov complexity on asymptotically Lifshitz rotating black branes}

\author{Mois\'es Bravo-Gaete$^{1}$}
\email{mbravo@ucm.cl,mchernicoff@ciencias.unam.mx,b.mellado.avila@gmail.com,juoliva@udec.cl,guadalupe.olmos@fisica.uaz.edu.mx}
\author{Mariano Chernicoff$^{2}$}
\author{Benjam\'in Mellado$^{3}$}
\author{Julio Oliva$^{3}$}
\author{Mar\'ia  Guadalupe Olmos Andrade$^{4}$}
\affiliation{
$^1$Departamento de Matem\'atica, F\'isica y Estad\'istica, Facultad de Ciencias
B\'asicas, Universidad Cat\'olica del Maule, Casilla 617, Talca, Chile.}
\affiliation{$^2$Departamento de F\'isica, Facultad de Ciencias,
UNAM,
A.P. 70-542, CDMX 04510, M\'exico.}
\affiliation{$^3$ Departamento de F\'isica, UdeC, Casilla, 160-C, Concepci\'on, Chile}
\affiliation{$^4$ Universidad Aut\'onoma de Zacatecas, Unidad Acad\'emica de F\'isica.}
\begin{abstract}
We study two types of holographic probes on a five-dimensional, asymptotically Lifshitz, rotating black hole. This geometry solves the field equations of Einstein gravity coupled to a dilaton, two axionic scalars, and a Maxwell doublet, supplemented by two vector doublets that enter the action through a parity-odd, Chern-Simons-like combination, as recently found in \cite{Bravo-Gaete:2026bdq}. This spacetime provides a concrete bottom-up arena to explore the imprints of rotation on probes in non-relativistic holography. First, we compute the holographic static quark-antiquark potential following the standard prescription that relates this boundary observable to the on shell-action of a string configuration whose endpoints are anchored at the boundary. We find that the screening length decreases both when the angular momentum is increased at fixed temperature and when the temperature is increased at fixed angular momentum. The former reflects centrifugal effects on the $q\bar{q}$ pair, whereas the latter stems from enhanced thermal fluctuations, which weaken correlations at a given separation. Second, we compute the growth rate of the Krylov complexity, using the holographic prescription that relates this observable to the momentum of a massive particle probe falling into the IR. In particular, we find that, at a given instant, the complexity grows faster as the dynamical exponent $z$ is increased.
\end{abstract}
\maketitle

\section{Introduction}\label{sec:introduction}

The holographic correspondence identifies certain quantum gauge theories with string theories in higher-dimensional curved spacetimes \cite{Maldacena:1997re,Gubser:1998bc,Witten:1998qj}. Its best-understood example relates four-dimensional $\mathcal{N}=4$ super Yang–Mills theory to type IIB string theory on $AdS_{5}\times S^{5}$. In an appropriate large-$N$, strong ’t Hooft coupling limit, the gauge theory is strongly coupled while its string dual is weakly coupled and weakly curved, allowing many observables to be computed using classical gravity and string probes.

Strongly coupled nonrelativistic field theories also present an important challenge, especially when describing quantum critical matter, where collective fluctuations can invalidate a quasiparticle description (e.g.\cite{Sachdev:2011fcc}). Such systems may exhibit anisotropic scaling, $t\to\lambda^z t$ and $\mathbf{x}\to\lambda\mathbf{x}$, with a dynamical exponent $z$ relating temporal and spatial correlation scales. In holography, Lifshitz geometries realize this symmetry as a bulk isometry when accompanied by the radial transformation $r\to\lambda^{-1}r$. They therefore provide candidate gravitational descriptions of strongly coupled nonrelativistic theories with this scaling, extending holography beyond anti-de Sitter asymptotics (see \cite{Kachru:2008yh} and references therein and thereof).

Within this framework, several observables characterize the equilibrium properties and dynamical response of strongly correlated matter. Black-hole thermodynamics determines quantities such as entropy and specific heat, while fluctuations of bulk fields yield correlation functions and spectral densities, together with transport coefficients such as electrical and thermal conductivities. Introducing suitable charged matter also permits the study of condensates and superconducting phase transitions. These computations provide concrete models of quantum critical transport and collective ordering in regimes where conventional descriptions based on weakly interacting quasiparticles become inadequate.

In this work, we consider the exact five-dimensional charged and rotating asymptotically Lifshitz black hole constructed in \cite{Bravo-Gaete:2026bdq}, with $2<z<3$. The solution is supported by a dilaton field, gauge fields, and axionic fields, supplemented by generalized Chern–Simons interactions. Its natural holographic interpretation is a finite-temperature state of a strongly coupled nonrelativistic theory in three spatial dimensions, carrying electric charge density and momentum along a compact spatial direction, interpreted as angular momentum. The axionic profiles supply spatially dependent scalar sources, while the radial dilaton profile encodes a scale-dependent coupling. The geometry thus defines an effective holographic model whose field-theory interpretation depends on its supporting matter and corresponding boundary conditions which should be consistent with Lifshitz asymptotics.

This background allows us to investigate how anisotropic scaling, temperature, and rotation jointly affect interactions and excitation dynamics. Since Lifshitz symmetry contains no boost symmetry, these effects cannot generally be inferred from a static configuration by changing the reference frame. Moreover, the rotation parameter controls the electric and axionic charges of this family, so its variation explores a correlated change of the state and supporting fields. We examine this setting through two complementary probes: the static quark–antiquark potential and the growth rate of spread Krylov complexity, which address spatial interactions and quantum-state evolution, respectively.

The quark–antiquark potential probes the interaction between external heavy color sources and its screening by the medium. Holographically, the $q\bar{q}$-pair is represented by a U-shaped string whose endpoints are anchored at the asymptotic boundary and whose worldsheet extremizes the Nambu–Goto action \cite{Rey:1998ik,Maldacena:1998im}. Here, the endpoints are separated along the compact rotation direction, and the corresponding potential is regularized by subtracting the energy of two straight strings extending down to the ergoregion. The spatial separation at which the connected pair is no longer energetically favored over the two disconnected strings defines the screening length.

The spread Krylov complexity measures how an evolving quantum state spreads over the orthonormal basis generated by successive action of its Hamiltonian. A relation between its growth rate and the proper radial momentum of an infalling massive particle has been established for locally excited states in two-dimensional conformal field theories with an $AdS_3$ description. We adopt its proposed extension to Lifshitz backgrounds \cite{BitaghsirFadafan:2026lek} (see also \cite{Vasli:2023syq} and \cite{Imani:2025etp}), representing a localized boundary excitation by a massive particle released near the asymptotic boundary. We solve its trajectory in the rotating geometry, including the angular motion induced by frame dragging, and evaluate the radial momentum using the reduced dynamics at fixed conserved angular momentum by considering the corresponding Routhian as in \cite{chatzis:2026}. Its dependence on boundary time then determines the proposed growth rate of spread Krylov complexity.

The paper is organized as follows: In Section~\ref{sec:background} we review the Lifshitz backgrounds, from the static black brane to the rotating solution, together with their thermodynamics.
In Section~\ref{sec:potential} we compute the holographic quark-antiquark
potential on the stationary background and study its
dependence on the rotation parameter and on the temperature. In
Section~\ref{sec:krylov} we compute the growth rate of the Krylov
complexity from the geodesic motion of a massive particle, and analyze its behavior as the charges controlling the solution are varied. Section \ref{sec:conc} contains some conclusions and comments.

\section{Lifshitz Solutions}\label{sec:background}
This section introduces the gravitational backgrounds underlying our holographic calculations. We begin with the five-dimensional Lifshitz geometry originally constructed in \cite{Bravo-Gaete:2026bdq}, explaining how the dynamical exponent $z$ determines its anisotropic scaling symmetry and how its isometries differ from those of anti-de Sitter spacetime. We then review the introduction of finite temperature and electric charge through static black-brane solutions, before presenting the charged rotating background considered in this work. We describe the matter fields supporting this solution, identify its event horizon and ergosurface, and obtain its Hawking temperature and entropy. Finally, we explain its non-rotating limit, establishing the geometric setting for the string and particle probes studied in the following sections.

Lifshitz geometries \cite{Kachru:2008yh} provide a gravitational realization of anisotropic scaling between time and space. Their defining parameter is the dynamical exponent $z$, which specifies how temporal scales change relative to spatial scales: under a rescaling of spatial distances by $\lambda$ and time scales by $\lambda^z$. The simplest five-dimensional realization, with homogeneous and isotropic spatial sections, is
\begin{equation}
ds^2=-\left(\frac{r}{\ell}\right)^{2z}dt^{2}
+\frac{\ell^{2}}{r^{2}}dr^{2}
+r^{2}\delta_{ij}dx^{i}dx^{j},
\qquad i,j=1,2,3.
\end{equation}
Here $\ell$ sets a curvature scale, and the asymptotic boundary lies at $r\rightarrow\infty$. We take the spatial coordinates $x^{i}$ to be dimensionless, consistently with the black-brane metrics introduced below. Equivalently, defining $X^{i}=\ell x^{i}$ puts the spatial line element in the form $r^{2}d\mathbf X^{2}/\ell^{2}$.
This metric is invariant under time translations, spatial translations, spatial rotations, and the anisotropic dilatation 
\begin{equation}
t\rightarrow\lambda^{z}t,
\qquad
x^{i}\rightarrow\lambda x^{i},
\qquad
r\rightarrow\lambda^{-1}r.
\end{equation}
The radial transformation compensates the rescaling of the boundary coordinates and leaves the line element unchanged. For generic $z>1$, these transformations, together with spatial rotations, constitute the Lifshitz isometry group. In particular, the geometry admits neither Lorentz nor Galilean boosts, and therefore selects a preferred time direction. When $z=1$, the metric becomes $AdS_{5}$ in Poincaré coordinates, and its local isometry algebra enlarges.
For $z\neq1$, the Lifshitz metric requires matter beyond a cosmological constant to solve the Einstein equations. Below we review the model that admits the rotating background on which we compute the holographic probes. The static, finite-temperature representative of such family can be described by \cite{Bravo-Gaete:2026bdq}

\begin{equation}
ds^2=-\Big{(}\frac{r}{\ell}\Big{)}^{2z} h(r)dt^2+\frac{{\ell}^2dr^2}{r^2h(r)}+r^2\delta_{ij}dx^idx^j,
\end{equation}
with
\begin{equation}
h(r)=1-\frac{M{\ell}^{z+3}}{r^{z+3}}\ .
\end{equation}
The event horizon $r_{+}$ is the largest positive zero of $h(r)$. 
The corresponding Hawking temperature is given by  
\begin{equation}
T_0=\frac{(z+3)r_{+}^{z}}{4\pi{\ell}^{z+1}}\ ,
\end{equation}
while the Bekenstein-Hawking entropy is
\begin{equation}
S_0=\frac{2\pi\sigma r_{+}^{3}}{\kappa}.
\end{equation}
At fixed $z$ and $\ell$, this branch satisfies $S_{0}/\sigma\propto T_{0}^{3/z}$, and taking $r_{+}\rightarrow0$ recovers the Lifshitz background \cite{Kachru:2008yh}.
Let us notice that a finite-temperature solution introduces a scale and is therefore not invariant under the Lifshitz dilatation at fixed parameters. Instead, scaling relates solutions with different horizon radii and temperatures. The static charged family provides a reference for introducing finite temperature and electric charge; the rotating solution considered below requires a different supporting matter sector.

Rotation introduces angular momentum and frame dragging, providing additional structure for the motion of particles and extended probes. Because boosts are absent from the Lifshitz isometry group, their effects cannot generally be obtained by applying an asymptotic boost symmetry to a static configuration. We therefore consider the exact charged rotating solution constructed in \cite{Bravo-Gaete:2026bdq}.

This new solution is an extremum of the theory containing the metric, a dilaton $\phi$, two abelian gauge fields $A^{(I)}$, two axionic scalars $\psi^{(I)}$, and two auxiliary one-form doublets $B^{(I)}$ and $C^{(I)}$, with $I=1,2$. Defining
\begin{equation}
F^{(I)}=dA^{(I)},
\qquad
H^{(I)}=dB^{(I)},
\qquad
K^{(I)}=dC^{(I)},
\end{equation}
the action is given by
\begin{equation}
\begin{aligned}
I=
\frac{1}{2\kappa}
\int d^{5}x,\sqrt{-g}
\left[
R-2\Lambda-\frac12(\partial\phi)^{2}
-\sum_{I=1}^{2}
\left(
e^{\alpha_I\phi}(F^{(I)})^{2}
+\frac12e^{\eta\phi}(\partial\psi^{(I)})^{2}
\right)
\right]
\\
+
\frac{2}{\kappa}
\sum_{I=1}^{2}\lambda_I
\int A^{(I)}\wedge H^{(I)}\wedge K^{(I)}.
\end{aligned}
\end{equation}
The generalized Chern-Simons interactions are metric independent and therefore make no direct contribution to the stress tensor. Their variation nevertheless modify the Maxwell equations, supplying the currents required by the rotating configuration.
The couplings and cosmological constant obey
\begin{equation}
k=\sqrt{6(z-1)}\ ,
\qquad
\alpha_1=\eta=-\frac6k \ ,
\qquad
\alpha_2=\frac2k \ ,
\qquad
\Lambda=-\frac{(z+3)(z+2)}{2\ell^{2}}\ .
\end{equation}
In particular, the second Maxwell coupling differs from that of the standard static charged construction, where its dilaton exponent is $2(z-1)/k$. The two charged families consequently belong to different matter models, although they share the static branch without the additional electric charge. Finally, the metric can be written as 
\begin{align}\label{metric}
\dd s^2=&-n(r)^{2}f(r)dt^{2}
+\frac{dr^{2}}{f(r)}
+r^{2}\bigl(dx^{1}+N_\phi(r)dt\bigr)^{2}
+r^{2}\bigl[(dx^{2})^{2}+(dx^{3})^{2}\bigr]\ ,\\
=&-F(r)\dd t^2+G(r)\dd r^2
       +r^2\bigl(\dd x+N_\phi(r)\dd t\bigr)^2
       +r^2(\dd x_2^2+\dd x_3^2)\ ,
\end{align}
where we have written the spacetime in two different ways, which will be useful in what follows, and
\begin{equation}
f(r)=
\frac{r^{2}}{\ell^{2}}
-\frac{M\ell^{z+1}}{r^{z+1}}
+\frac{(5-z)\ell^{2z-2}}{2(z-2)}
\frac{J^{2}}{r^{6}}\ ,
\end{equation}

\begin{equation}
n(r)=
\left(\frac r\ell\right)^{z-1}\ ,
\qquad
N_\phi(r)=\frac{J}{r^{5-z}}\ .
\end{equation}
The integration constants $M$ and $J$ control the mass and rotation, respectively. We take $x^{1}$ to be compact, with period $\sigma_1$, so that momentum along this direction is interpreted as angular momentum. Notice that there is no frame rotation at infinity since $g_{tx^1}$ vanishes as $r\rightarrow\infty$ for the cases of interest $2<z<3$.
For completeness, we include here the dilaton, Maxwell potentials, and axionic profiles that are given by
\begin{equation}
e^{\phi}=\mu r^{k}\ ,
\end{equation}
\begin{equation}
A^{(1)}=
\sqrt{\frac{2(z-1)}{z+3}}
\frac{\mu^{3/k}}{2\ell^{z}}
\left(r^{z+3}-r_{+}^{z+3}\right)dt\ ,
\end{equation}
\begin{equation}
A^{(2)}=
\sqrt{\frac{3(3-z)}{4(z-2)}}
\frac{J}{\mu^{1/k}}
\left(r^{z-5}-r_{+}^{z-5}\right)dt\ ,
\end{equation}
\begin{equation}
\psi^{(I)}=
(5-z)\ell^{z-1}\mu^{3/k}Jx^{I+1},
\qquad I=1,2\ .
\end{equation}
Here $\mu>0$ fixes the scalar normalization, and the additive constants in the gauge potentials have been chosen in order that they vanish at the horizon. The auxiliary one-forms are chosen as in \cite{Bravo-Gaete:2026bdq} to satisfy the modified Maxwell equations. These expressions show that the additional electric field and the axionic gradients are tied to $J$; varying rotation therefore changes the accompanying matter configuration as well. The event horizon is located at the largest positive root of $f(r_{+})$, its Hawking temperature is given by
\begin{equation}
T=\frac{n(r_{+})f'(r_{+})}{4\pi}=\frac{(z+3)r_{+}^{z}}{4\pi\ell^{z+1}}-\frac{(5-z)^{2}J^{2}\ell^{z-1}}{8\pi(z-2)r_{+}^{8-z}}\ ,
\end{equation}
while the entropy reads
\begin{equation}
S=\frac{2\pi\sigma r_{+}^{3}}{\kappa}\ .
\end{equation}
Although this last expression contains no explicit dependence on $J$, rotation affects the relation between $r_{+}$ and the thermodynamic parameters. The event horizon must also be distinguished from the so-called stationary-limit surface, or ergosurface,
defined by
\begin{equation}
g_{tt}(r_e)=-n(r_e)^{2}f(r_e)
+{r_e}^{2}N_\phi(r_e)^{2}
=0\ .
\end{equation}
For $J\neq0$, $g_{tt}(r_{+})=r_{+}^{2}N_\phi(r_{+})^{2}>0$, whereas $g_{tt}<0$ asymptotically. Consequently, an ergosurface lies outside the event horizon. Between these two surfaces, trajectories at fixed spatial coordinates cannot remain timelike. This distinction will be relevant below, when considering both stationary string configurations and the frame dragging of infalling particles.

Setting $J = 0$ removes the shift, the second electric field, and the axionic gradients. The metric
function $f(r)$ then reduces to
\begin{equation}
f(r)=\frac{r^{2}}{\ell^{2}}
\left(
1-\frac{M\ell^{z+3}}{r^{z+3}}
\right),
\end{equation}
which reproduces $h_0(r)$, $T_0$, and $S_0$ after identifying $r_{+}^{z+3}=M\ell^{z+3}$. Thus, the nonrotating limit is the static branch without the additional electric charge. 

Holographically, the rotating background provides an effective description of a finite-temperature nonrelativistic state with electric charge and momentum along a compact spatial direction. The axionic profiles specify additional spatially dependent sources, while the running dilaton encodes a scale-dependent coupling. This setting allows us to study how the combined thermal, rotational, and matter contributions influence external source interactions and the evolution of the state of the dual field theory in the Krylov basis. 
\section{HOLOGRAPHIC QUARK-ANTIQUARK POTENTIAL}
\label{sec:potential}
According to the holographic correspondence, a static infinitely massive quark-antiquark pair is dual to a U-shape string, whose endpoints are anchored at the asymptotic boundary. To be more precise, the two endpoints represent the quark and antiquark, while the U-shape string extending into the bulk codifies information about the gluonic fields. 

The static quark–antiquark potential characterizes the interaction between these two external, infinitely massive color sources held at a fixed separation $L$. From the gauge theory viewpoint, this interaction is encoded in a rectangular temporal Wilson loop \cite{Rey:1998ik,Maldacena:1998im}. Its two long sides follow the trajectories of the quark and antiquark, while its spatial extent specifies their separation. The dependence of the expectation value on a long observation time determines the energy of the pair; removing the contributions associated with the individual color sources then isolates their interaction energy.

In this Section, we adopt this prescription for the rotating, asymptotically Lifshitz geometry introduced in Section II. Notice that this prescription has already been established in the context of asymptotically Lifshitz, static backgrounds \cite{BitaghsirFadafan:2026lek,Zhang:2016jns}. The infinitely heavy external sources are fixed at the boundary and separated along the compact direction. The dual interpretation of the two unbound static color sources corresponds to two disconnected straight strings extending from the boundary to a  probe brane located at $r=r_e$. With this prescription, we compute the static quark–antiquark potential and examine how temperature and rotation affect its dependence on the separation and on the associated screening length.

The Nambu-Goto string action is given by
\begin{equation}
S_{\mathrm{NG}}=-\frac{1}{2\pi\alpha'}\int d\sigma d\tau\sqrt{-\det g_{ab}} \ ,
\end{equation}
where $T_s=\frac{1}{2\pi\alpha'}$ is the string tension and $g_{ab}$ is the induced metric on the string worldsheet. Choosing the static gauge, i.e. $\tau=t$, $\sigma=x$, the profile of the U-shape string can be described by the embedding $X^\mu(t,x)=\bigl(t,r(x),x,x_2,x_3\bigr)$ on the spacetime \eqref{metric}. The corresponding boundary conditions for $r(x)$ are 
\begin{equation}
 r(\pm L/2)=r_{UV},\quad r(0)=r_{\text{ret}}\quad \text{and} \quad r'(0)=0\ ,
 \label{eq:embedding}
\end{equation}
where $r_{UV}$ is a radial UV cutoff, $r_{\text{ret}}$ is the turning point of the U-shape string in the bulk and $r'=\dd r/\dd x$. We work on the Euclidean saddle obtained by the Wick rotation $t\rightarrow -it_E$, without any further rotation on the integration constants therefore, the saddle becomes complex. 

The relevant parameters for our case are (i) those related to the asymptotic geometry $(z,\ell)$, (ii) the black hole parameters $(M,J)$ as well as $r_+=r_+(M,J)$,  and (iii) those of the string configuration $(r_\text{ret},\Delta\tau)$ the latter being the length of the time lapse, which is taken to infinity in order to match the infinite, rectangular loop in the boundary theory. The regulator $r_\text{UV}$ is eventually taken to infinity as well. The quark-antiquark pair separation $L$ can be written in terms of these variables and it is given by the integral

\begin{equation}
   L=2\int_{r_\text{ret}}^{r_\text{UV}} dr \sqrt{\frac{F(r_\text{ret})r_\text{ret}^2G(F-J^2r^{2z-8})}{Fr^2(Fr^2-F(r_\text{ret})r_\text{ret}^2)}}\ .
\end{equation}
Notice that $r_\text{ret}$, i.e. the location of the U-shape turning point cannot be located between the horizon $r=r_+$ and the ergosphere $r=r_\text{ergo}$, where the combination $(F-J^2r^{2z-8})$ becomes negative, meaning that the separation of the quark and antiquark in this static ansatz will not be well defined. Therefore, we restrict to the case in which $r_\text{ret}\geq r_\text{ergo}$. There is a natural interpretation for this fact: it is not possible to have a static worldsheet if the string penetrates into the ergoregion. 

\begin{figure}[htbp]
\centering
\resizebox{0.6\linewidth}{!}{\includegraphics{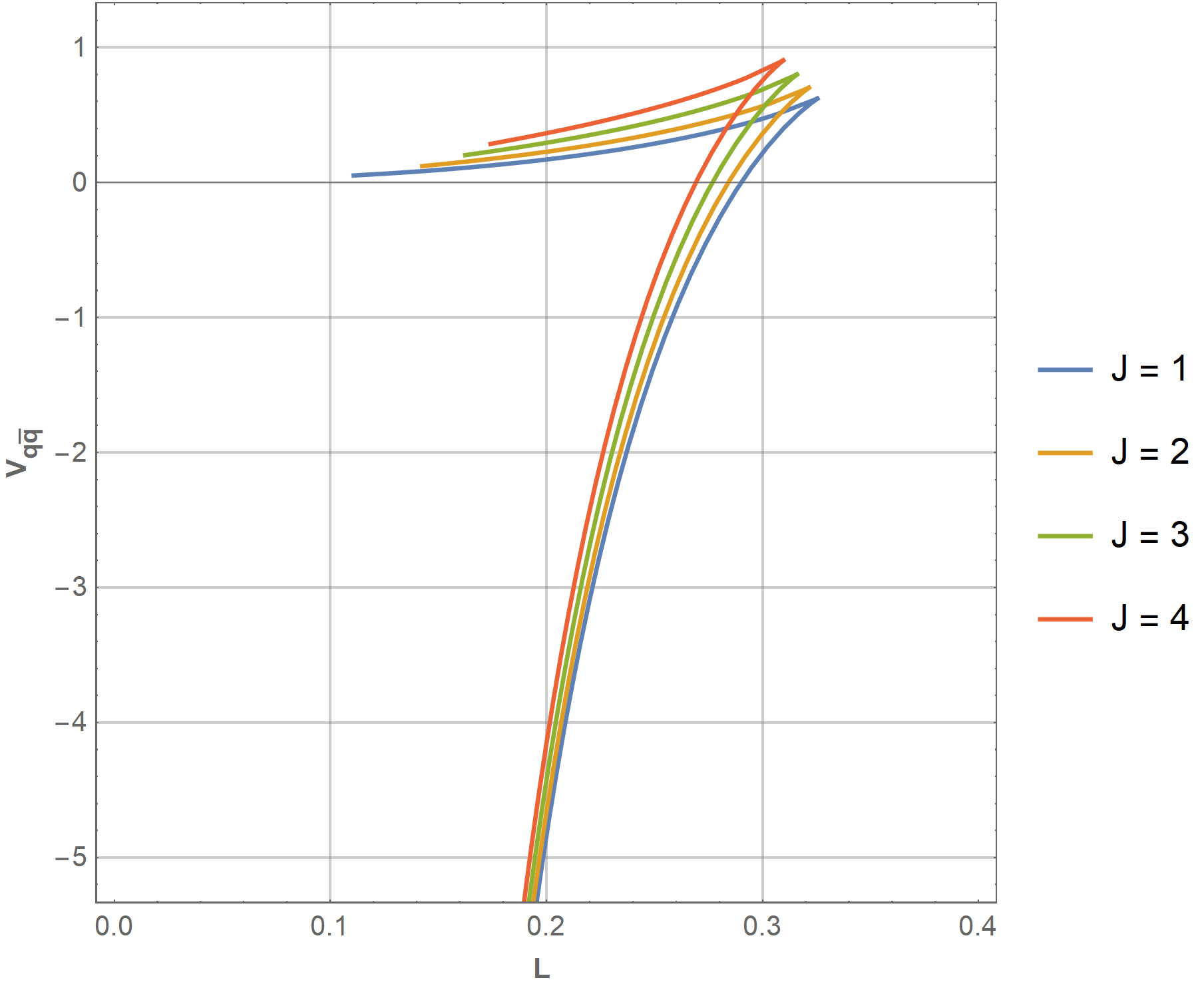}}
\caption{Quark-antiquark potential $V=\Vb$ versus separation $L$ at fixed $T=2$, for $J=1,2,3,4$. Here we have set the anisotropic exponentto $z=5/2$.}
\label{fig:potential-J}
\end{figure}

\begin{figure}[htbp]
\centering
\resizebox{0.6\linewidth}{!}{\includegraphics{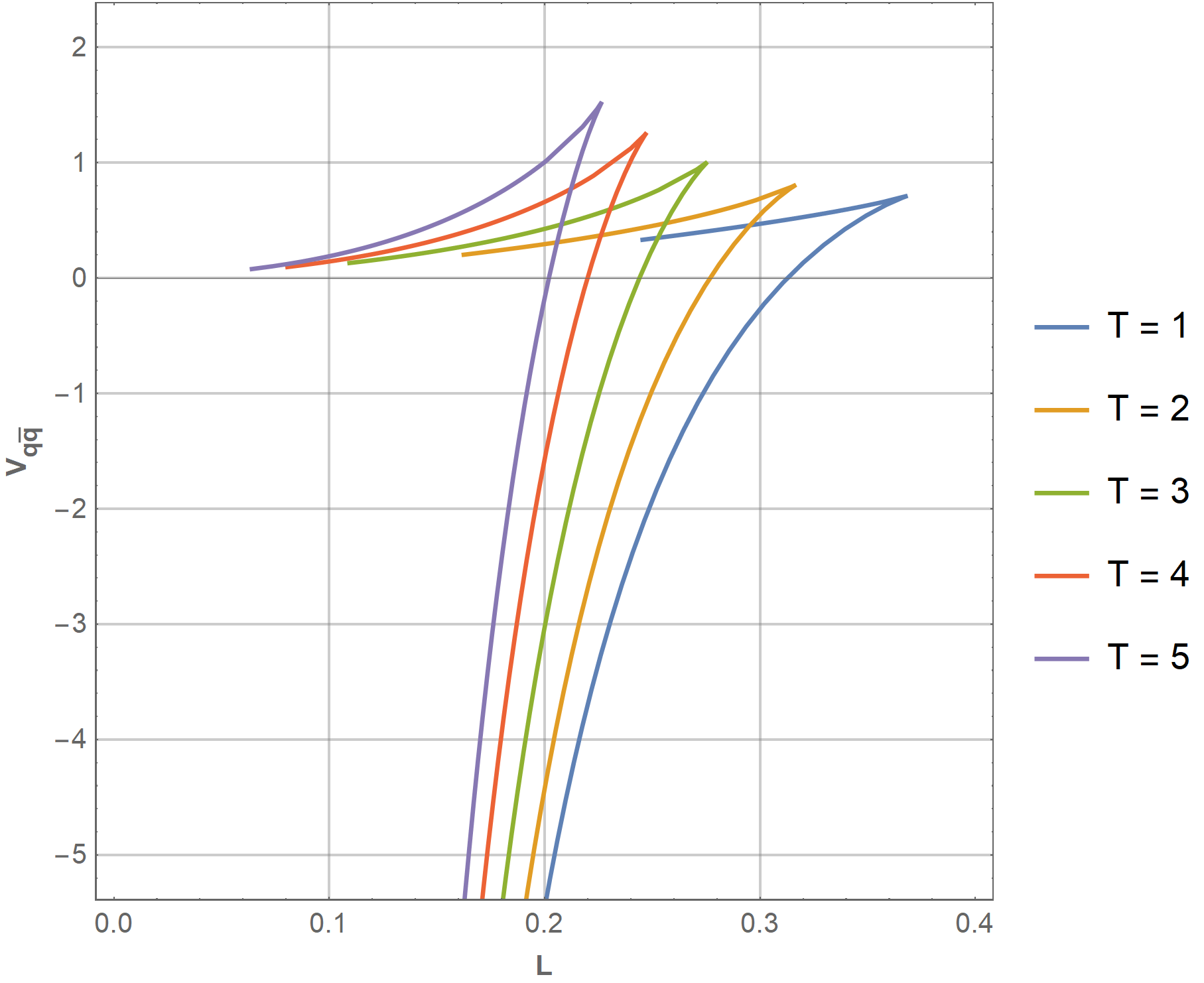}}
\caption{Quark-antiquark potential $V=\Vb$ versus $L$ at fixed $J=3$, for $T=1,2,3,4,5$. Here we have set the anisotropic exponent to $z=5/2$.}
\label{fig:potential-T}
\end{figure}

The on-shell Euclidean Nambu-Goto action diverges in the asymptotically Lifshitz background, therefore we regularize such infinity by substracting the action on two disconnected strings that fall straight from the boundary up the ergosphere $r=r_\text{ergo}$ bringing in a probe brane on the ergosurface where the ending points of the straight strings are anchored. The regularized Nambu-Goto action then reads
\begin{align*}
    S^\text{reg}=&\frac{\Delta\tau}{\pi\alpha'} \int_{r_\text{ret}}^{r_\text{UV}}dr\sqrt{G\left(F-J^2r^{2z-8}\right)}\left[\sqrt{\frac{Fr^2}{Fr^2-r_\text{ret}^2F(r_\text{ret})}}-1\right]\\
    &-\frac{\Delta\tau}{\pi\alpha'} \int_{r_\text{ergo}}^{r_\text{ret}}dr\sqrt{G\left(F-J^2r^{2z-8}\right)}\ ,
\end{align*}
where again we have assumed that the turning point does not penetrate the ergoregion.
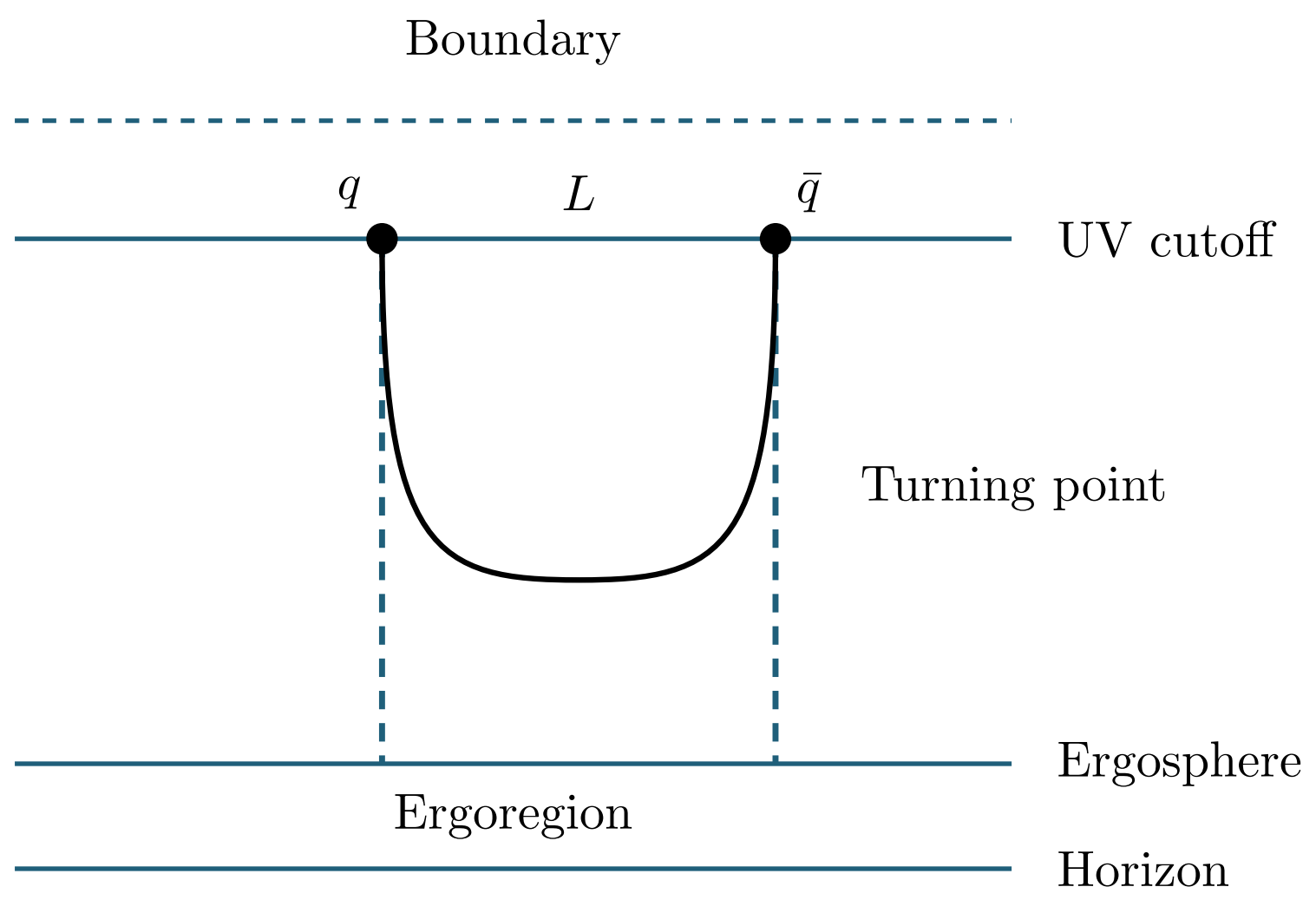
\begin{figure}[h]
  \centering
\begin{tikzpicture}[line width=1pt, bluegray]
 
  Horizontal lines
  \draw[dashed]  (0,5.4) -- (7.6,5.4);   
  \draw          (0,4.5) -- (7.6,4.5);   
  \draw          (0,0.5) -- (7.6,0.5);   
  \draw          (0,-0.3) -- (7.6,-0.3); 
 
  Vertical dashed lines (clearly visible)
  \draw[line width=1.3pt, dash pattern=on 4pt off 3pt]
    (2.8,4.5) -- (2.8,0.5);
  \draw[line width=1.3pt, dash pattern=on 4pt off 3pt]
    (5.8,4.5) -- (5.8,0.5);
 
  Smooth U-shaped string (tangent-continuous at the bottom)
  \draw[black, line width=1.2pt]
    (2.8,4.5) .. controls (2.8,2.1) and (3.2,1.9) .. (4.3,1.9)
              .. controls (5.4,1.9) and (5.8,2.1) .. (5.8,4.5);
 
  Quark and antiquark endpoints
  \fill[black] (2.8,4.5) circle (0.12);
  \fill[black] (5.8,4.5) circle (0.12);
 
  Labels
  \node[black] at (2.55,4.85) {$q$};
  \node[black] at (6.05,4.85) {$\bar{q}$};
 \node[black] at (4.3,4.85)  {$L$};
 \node[black] at (3.8,6.0)   {Boundary};
 
 \node[black, anchor=west] at (7.8,4.5)  {UV cutoff};
 \node[black, anchor=west] at (6.3,2.6)  {Turning point};
 \node[black, anchor=west] at (7.8,0.5)  {Ergosphere};
 \node[black, anchor=west] at (7.8,-0.3) {Horizon};
 \node[black] at (3.8,0.1) {Ergoregion};
 
\end{tikzpicture}
\caption{String configuration leading to the regularized Nambu-Goto action. The dashed vertical lines represent the configuration being substracted in order to regularize the action, and extract the potential energy due to the interaction in the $q\bar{q}$-pair.}
\end{figure}
On the rotating background, the integration has to be performed numerically. The potential depends on the temperature and the angular momentum of the black hole background. In \ref{fig:potential-J} we have fixed the temperature to a reference value of $T=2$, and have plotted the $q\bar{q}$ potential as a function of their separation for different values of the angular momentum. The screening length is seen to decrease with the angular momentum, which suggests that there is an inherited centrifugal effect on the boundary $q\bar{q}$ pair. A similar behavior has been seen in the asymptotically AdS case \cite{Golubtsova:KerrAdS}. 
On the other hand in \ref{fig:potential-T} we fixed the angular momentum to a reference value $J=3$, and plotted the potential for different values of the temperature. Again, the screening length decreases with the temperature, which is natural since thermal fluctuations suppress correlation at a given distance.
From these plots, it is possible to see that for small separations, the potential scales as $L^{-z}$, as expected from the anisotropic scaling symmetry.

\begin{figure}[htbp]
\centering
\begin{minipage}[t]{0.49\linewidth}
\centering
\resizebox{\linewidth}{!}{\includegraphics{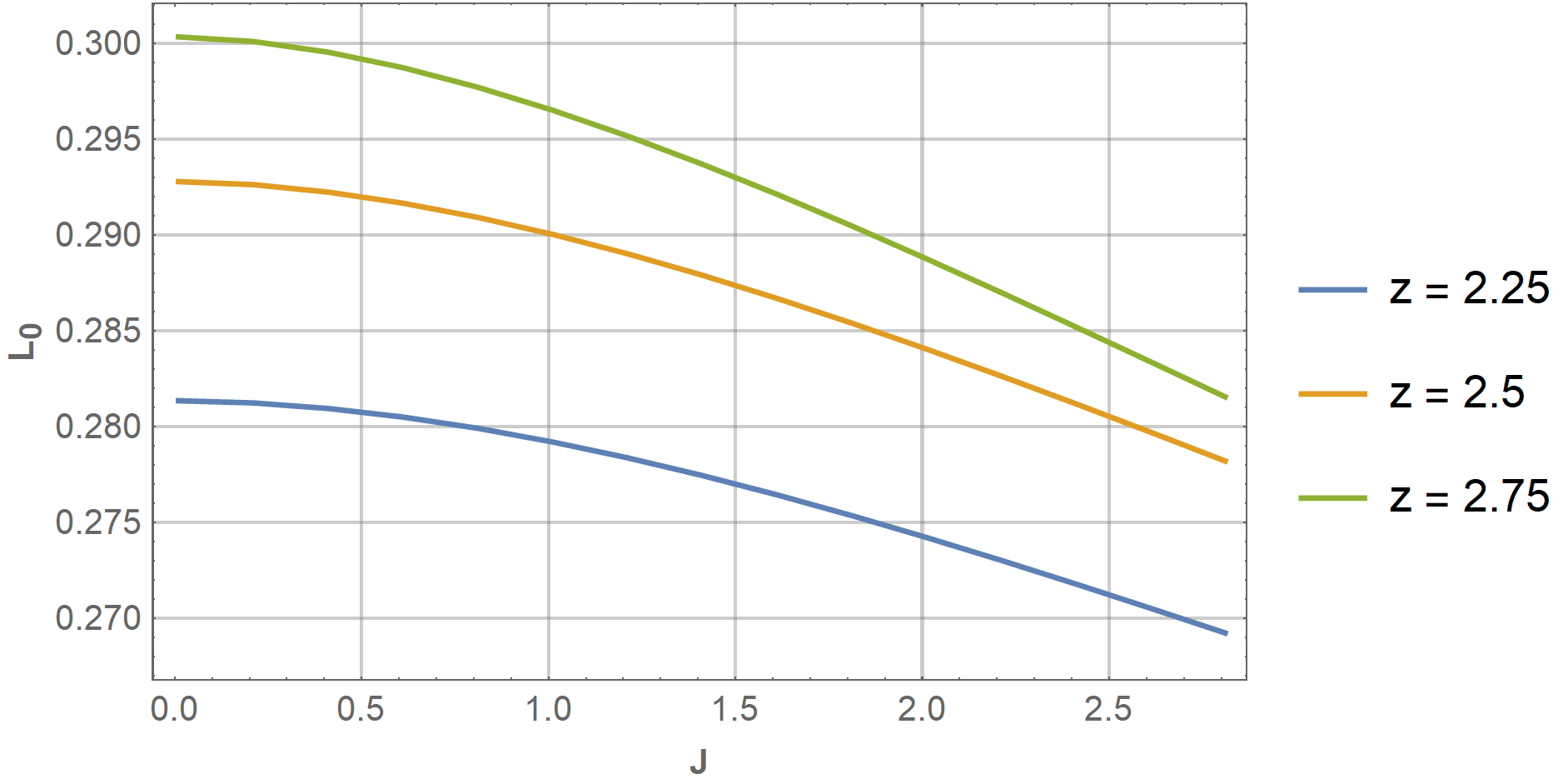}}\\[3pt]
\textnormal{(a) Fixed $T=2$.}
\end{minipage}\hfill
\begin{minipage}[t]{0.49\linewidth}
\centering
\resizebox{\linewidth}{!}{\includegraphics{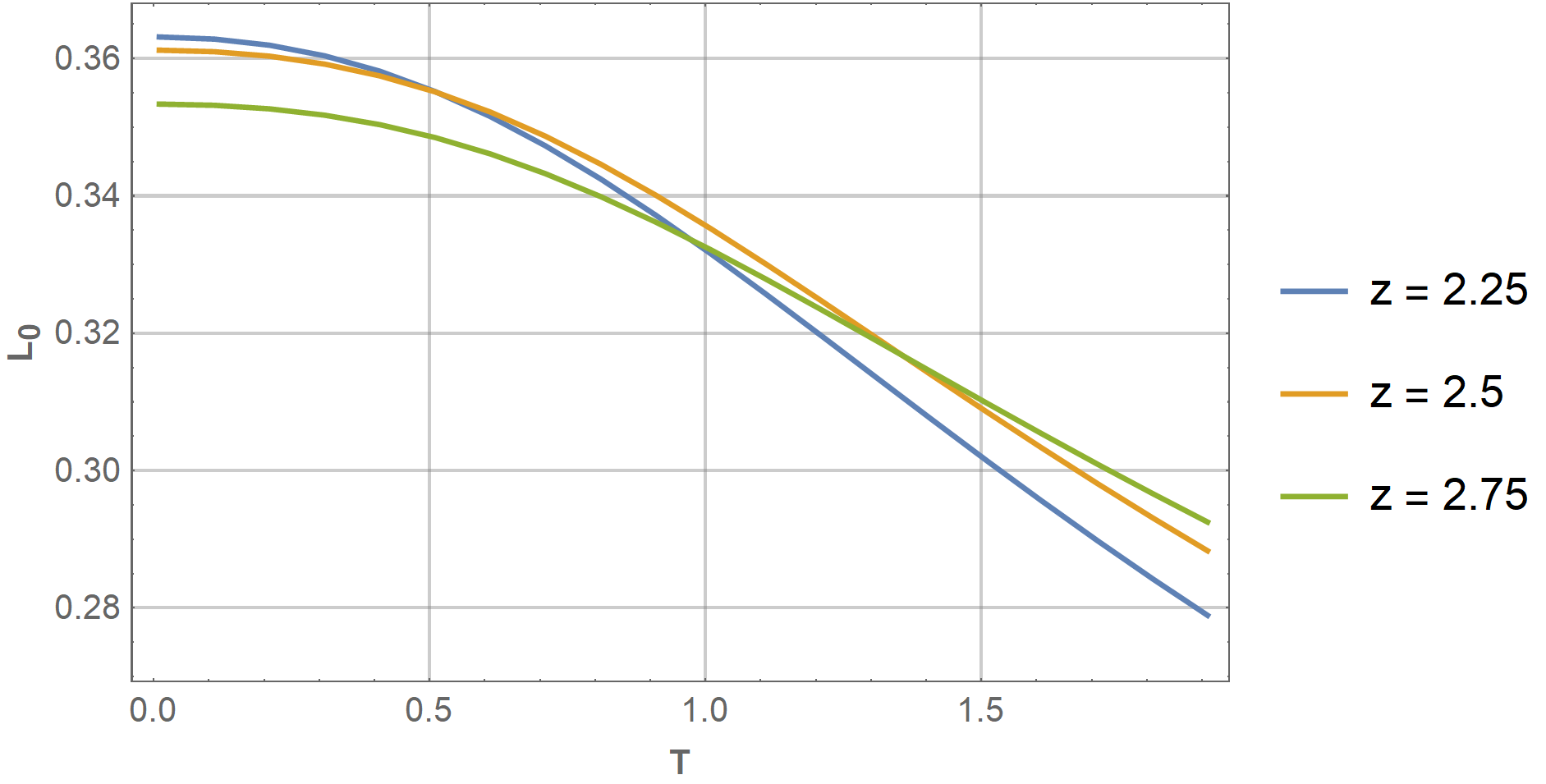}}\\[3pt]
\textnormal{(b) Fixed $J=2$.}
\end{minipage}
\caption{Screening length $L_0$ as a function of (a) the rotation parameter $J$ and (b) the temperature $T$ for different values of the dynamical exponent $z$.}
\label{fig:screening-scans}
\end{figure}

Finally, our setup permits to compute the effect of the Lifshitz exponent on the screening length. Figure \ref{fig:screening-scans} contains such information. It is seen that for a fixed temperature, increasing the dynamical exponent increases $z$ consistently increases the screening length $L_0$ for all the allowed values of $J$ explored. Nevertheless, once $J$ is fixed, for different regimes of the temperature the value of $L_0$ can increase or decrease with a rising value of $z$. It would be interesting to have a physical interpretation of this behavior.

\section{Growth rate of spread Krylov complexity}\label{sec:krylov}
The holographic dictionary has recently been extended to the computation of the growth rate of spread Krylov complexity \cite{Caputa:2021sib}. Such concept can also be studied in non-relativistic quantum theories \cite{Vasli:2023syq}, \cite{Imani:2025etp}, which may have a dual description allowing for the computation of this quantity from an asymptotically Lifshitz background \cite{BitaghsirFadafan:2026lek}. Here we follow this latter prescription in the context of our rotating, asymptotically Lifshitz black hole.

Consider a quantum system with a normalized initial state $\ket{\psi_0}$. Its evolution in the Schr\"odinger picture is
\begin{equation}
  \ket{\psi(t)} = e^{-iHt}\ket{\psi_0} = \sum_{k=0}^{\infty}\frac{(-it)^k}{k!}\,H^k\ket{\psi_0}.
\end{equation}
The evolved state therefore lies in the span of the set
$\{\ket{\psi_0}, H\ket{\psi_0}, H^2\ket{\psi_0}, \ldots\}$. Applying the Gram--Schmidt procedure to this set (equivalently, the Lanczos algorithm) yields an orthonormal set
$\{\ket{K_n}\}_{n=0}^{\infty}$ with $\ket{K_0}=\ket{\psi_0}$, known as the \emph{Krylov basis}. Although this set need not span the whole Hilbert space, it spans the subspace explored by the dynamics. Hence, at every time one can write
\begin{equation}
  \ket{\psi(t)} = \sum_{n=0}^{\infty}\phi_n(t)\,\ket{K_n},
  \qquad
  \sum_{n=0}^{\infty}|\phi_n(t)|^2 = 1,
\end{equation}
and in this basis the Hamiltonian is tridiagonal,
\begin{equation}
  H\ket{K_n} = a_n\ket{K_n} + b_{n+1}\ket{K_{n+1}} + b_n\ket{K_{n-1}},
\end{equation}
where $a_n$ and $b_n$ are the Lanczos coefficients. At $t=0$ only $\ket{K_0}$ is populated, and the Hamiltonian dynamics may progressively transfers probability to the higher Krylov states. Krylov complexity measures this spreading and is defined as the average position on the Krylov chain,
\begin{equation}
  C(t) = \sum_{n=0}^{\infty} n\,|\phi_n(t)|^2 .
\end{equation}
Remarkably, it has recently been shown that the growth rate of the spread
(Krylov) complexity of states excited by a local primary operator of large
dimension $\Delta\gg1$ in a two-dimensional CFT is given by the proper radial
momentum of a dual massive particle of mass $m\simeq\Delta$ falling in
AdS$_3$ \cite{Caputa:2024sux},
\begin{equation}\label{eq:ecPfromL}
  \dot{C}(t) = -\frac{1}{\epsilon}\,P_\rho(t),
  \qquad P_\rho=\frac{\partial L}{\partial\dot\rho},
\end{equation}
where $\epsilon$ is the regulator of the operator insertion and $\rho$ is the
proper radial distance (to the horizon, or to the center of global AdS). In
that setting the relation is derived from the $SL(2,\mathbb{R})$ symmetry
controlling the Krylov chain. This symmetry is absent in Lifshitz geometries,
so the extension to non-relativistic holography \cite{BitaghsirFadafan:2026lek} adopts the
same identification as a holographic proposal. Here we apply a similar prescription to a rotating, asymptotically Lifshitz black hole, which must be extended since purely radial motion is non-geodesic due to spacetime dragging even outside the ergosphere. In the case of the string probe, as far as the string with ends attached to the boundary do not penetrate the ergoregion, the tension of the string leads to the consistency of the static ansatz, nevertheless no such tension exist for a free massive particle falling into the interior, and the geodesic necessarily explores the angular directions. 

For the background \eqref{metric} we consider the geodesic of the probe of mass $m$ moving in the $(t,r,x)$ sector, with $x_2,x_3$ fixed, and falling from the UV cutoff $r_{\text{UV}}$ into the bulk. To this end, we define the proper-time interval $d\tau^2 = - ds^2$ and the parametrization 
\begin{align}
x^\mu(t)=
\big(t,r(t),x(t),x_{2},x_{3}\big),
\end{align}
we obtain 
\begin{align}
    d\tau = dt\sqrt{F(r) - G(r)\dot{r}^2 - r^2(\dot x + N_\phi(r))^2} \equiv dt\sqrt{B}
\end{align}
where the dot denotes $d/dt$ and the term under the square root is denoted by $B$ for brevity. The action $S=-m\int d\tau\equiv\int L\,dt$ has a Lagrangian
\begin{align}\label{eq:K-L}
    L = -m\sqrt{B}.
\end{align}

Since the Lagrangian does not depend on $t$ and $x$ is cyclic, the energy $E$ and the momentum $J_x$ along $x$ are conserved
\begin{align}\label{Jx}
    J_x := \frac{\partial L}{\partial \dot{x}}
    &= \frac{mr^2(\dot{x}+N_\phi(r))}{\sqrt{B}},
\end{align}

\begin{align}\label{energia}
    E &= \frac{\partial L}{\partial \dot{r}}\dot{r}
    + \frac{\partial L}{\partial \dot{x}}\dot{x}
    - L= \frac{mF(r)}{\sqrt{B}}-N_\phi(r)J_x.
\end{align}

It is convenient to introduce the functions $\mu(r)$ and $\mathcal{E}(r)$ by
\begin{equation}
  \mu(r)^{2}\equiv m^{2}+\frac{J_x^{2}}{r^{2}}\,,
  \label{eq:K-mu}
\end{equation}
and
\begin{equation}
  \mathcal E(r)\equiv E+N_\phi(r)J_x=\frac{m F(r)}{\sqrt{B}}=\frac{\mu F}{\sqrt{F-G\dot r^{2}}}\,.
  \label{eq:K-Ecal}
\end{equation}
In terms of these quantities
\begin{align}\label{eq:rdot}
\dot{r}^2
=\frac{F(r)}{G(r)}\left[
1-\mu^2 (r)\frac{F(r)}{\mathcal{E}^2}
\right],
\end{align}
and
\begin{align}\label{eq:K-xdot}
\dot{x} = \frac{J_x}{r^2} \frac{\sqrt{F(r) - G(r) \dot{r}^2}}{\mu(r)} - N_\phi(r).
\end{align}
Because $x$ is cyclic, $J_x$ is conserved along the whole trajectory. This is an obstacle if the proper momentum is built directly from the Lagrangian using a proper coordinate that includes the motion along $x$, as in \eqref{eq:ecPfromL}. In that case, the total spatial momentum contains a constant contribution from $J_x$, which is already present at $t=0$. As a result, $P_\rho(0) \sim J_x$, in contradiction with the requirement that the quantity identified with $\dot{C}$ must vanish at $t=0$.

The solution adopted in \cite{chatzis:2026} for this feature, was to eliminate the cyclic coordinate by considering the Routhian
\begin{align}
    \mathcal{R} &= \frac{\partial L}{\partial \dot x} \dot x - L,\\ 
    &=J_x \dot x (r,\dot r, J_x) - L(r,\dot r, J_x),\\
    &=\mu(r)\sqrt{F-G\dot r^{2}}-J_x\,N_\phi(r)\,,
\end{align}
where \eqref{eq:K-xdot} has been used to remove $\dot x$. This defines a purely radial problem in which $J_x$ is a parameter. Absorbing the $r$-dependent factor $\mu(r)$ into the proper radial coordinate,
\begin{align}
d\rho &= \mu(r)\sqrt{G(r)}\, dr,\\
\dot{\rho}^2 &\equiv \mu^2 (r) G (r)\dot{r}^2 ,
\end{align}
the proper momentum $P_\rho$ is given by
\begin{align}\label{eq:K-Prho}
P_\rho = - \frac{\partial \mathcal{R}}{\partial \dot{\rho}} = \frac{\dot{\rho}}{\sqrt{\mu^2 F(r) - \dot{\rho}^2}} = \frac{\sqrt{G (r)}\, \dot{r}}{\sqrt{F (r)- G(r) \dot{r}^2}}.
\end{align}
Since the probe falls inward, $\dot r<0$ and hence $P_\rho<0$, so that $\dot C\sim-P_\rho>0$.

The dynamics of the reduced problem follow from the Hamiltonian
\begin{equation}
  H=\dot r\,\frac{\partial \mathcal{R}}{\partial\dot r}-\mathcal{R}=-\frac{\mu F}{\sqrt{F-G\dot r^{2}}}+J_x\,N_\phi=-E\,,
\end{equation}
so that, defining the energy from the Routhian as $\mathcal H=-H$ as
in~\cite{chatzis:2026}, $\mathcal H=E$ and the radial equation coincides with the
one obtained from the full Lagrangian \eqref{eq:K-L} at fixed $J_x$: the
reduction changes the definition of the momentum but not the trayectory.

We release the probe from rest at the cutoff, $r_{\text{UV}}=1/\epsilon$, with
\begin{align}
    \dot r(0) = 0, \qquad \dot x(0)=0. 
\end{align}
In the asymptotic region $\epsilon \rightarrow 0$, the metric functions take the leading order form
\begin{align}
F(r_{\text{UV}}) \approx r_{\text{UV}}^{2z}=\epsilon^{-2z}, \qquad G(r_{\text{UV}})\approx r_{\text{UV}}^{-2}=\epsilon^{2},\qquad N_\phi(r_{\text{UV}}) = \frac{J}{r_{\text{UV}}^{5-z}}=J \epsilon^{5-z}.
\end{align}

With these conditions, the conserved quantities from \eqref{Jx} and \eqref{energia} become
\begin{align}
    J_x \approx \frac{m J}{r_{\text{UV}}^3}=mJ\epsilon^{3}, \qquad E \approx m r_{\text{UV}}^z=m\epsilon^{-z},
\end{align}
and therefore
\begin{align}
\mu^2(r) = m^2 + \frac{m^2 J^2 \epsilon^6}{r^2}, \qquad \mathcal{E}(r) = m \epsilon^{-z} + \frac{m J^2}{r^{5-z}} \epsilon^3.
\end{align}

Since $\dot r(0)=0$, the equation \eqref{eq:K-Prho} gives $P_\rho(0)=0$ directly; equivalently, $\mathcal E(r_{\text{UV}})=\mu\sqrt{F}\,|_{r_{\text{UV}}}$ and
\begin{equation}
  P_\rho^{2}=\frac{\mathcal E^{2}}{\mu^{2}F}-1
  \label{eq:K-Prho2}
\end{equation}
vanishes at $r_{\text{UV}}$, so $\dot C(0)\sim-P_\rho(0)=0$ as required.

The behavior at early times follows from the radial equation \eqref{eq:rdot}. Since the probe is released from rest at $r=r_\text{UV}$, expanding around this point gives $\dot r^2 \propto (r_\text{UV} -r)$, which integrates to $r_\text{UV} - r(t) \propto t^2$. In the same way, near the boundary $P_\rho \propto \sqrt{r_\text{UV}-r}$, so $P_\rho \propto t$. Using $\dot C_K \propto -P_\rho$ and integrating once, we confirm that the calculated complexity exhibits the expected behavior $C_K(t) \propto t^2 + \mathcal{O}(t^4)$ when $t \rightarrow 0$.

Finally, since the probe falls inward, $\dot {r} < 0$, and we can write $\dot{r}(r)=-\sqrt{\dot {r}^2 (r)}$, and using the change of variable $dt=-dr/\sqrt{\dot {r}^2 (r)}$ we can obtain an expression of $t(r)$ as an integral from $r_{\text{UV}}$ to $r$
\begin{align}\label{eq:integralt}
t(r) =\int_{r}^{r_{\text{UV}}} \frac{dr}{\sqrt{\dot r^2 (r)}} =\int_{r}^{r_{\text{UV}}} dr \, \frac{\mathcal{E}(r) \sqrt{G(r)}}{\sqrt{F(r) \left[ \mathcal{E}^2(r) - \mu^2(r) F(r) \right]}}.
\end{align}

Similarly, the growth rate of the complexity is governed by $\dot C(t) \propto -P_\rho$, so that 
\begin{align}\label{eq:integralC}
    C(r) \propto \int_{r}^{r_{\text{UV}}} dr \, \frac{\mathcal{E}(r) \sqrt{G(r)}}{\mu(r) F(r)}.   
\end{align}

The original proposal \cite{Caputa:2024sux} shown in \eqref{eq:ecPfromL}, is identifies $\dot C_K = -  P_\rho/\epsilon$. In Lifshitz background, the physical normalization is not fixed by the symmetry argument of AdS$_3$, and because the anisotropic scaling modifies the dimension of the regulator, for the numerical analysis we choose
\begin{equation}
    \dot C_K (t) = - \epsilon ^z P_\rho,
\end{equation}
which cancels the large redshift factor
$\epsilon^{-z}$ present in $P_\rho$ (since $E\simeq m\epsilon^{-z}$). 
With this choice, we now proceed to the analysis of the results.

\begin{figure}[htbp]
\centering
\begin{minipage}[t]{0.49\linewidth}
\centering
\resizebox{\linewidth}{!}{\includegraphics{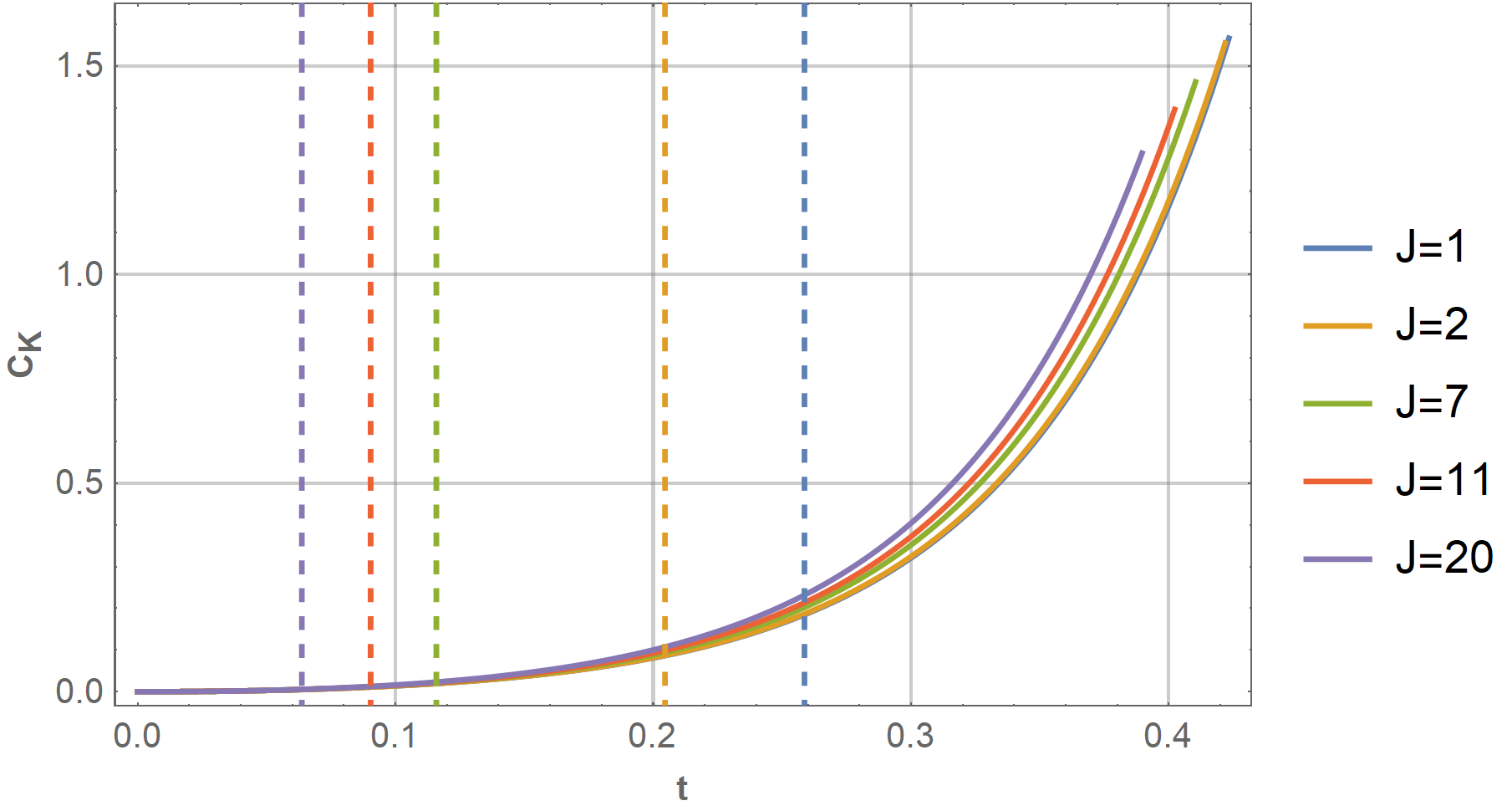}}\\[3pt]
\textnormal{(a) Behavior at early times.}
\end{minipage}\hfill
\begin{minipage}[t]{0.49\linewidth}
\centering
\resizebox{\linewidth}{!}{\includegraphics{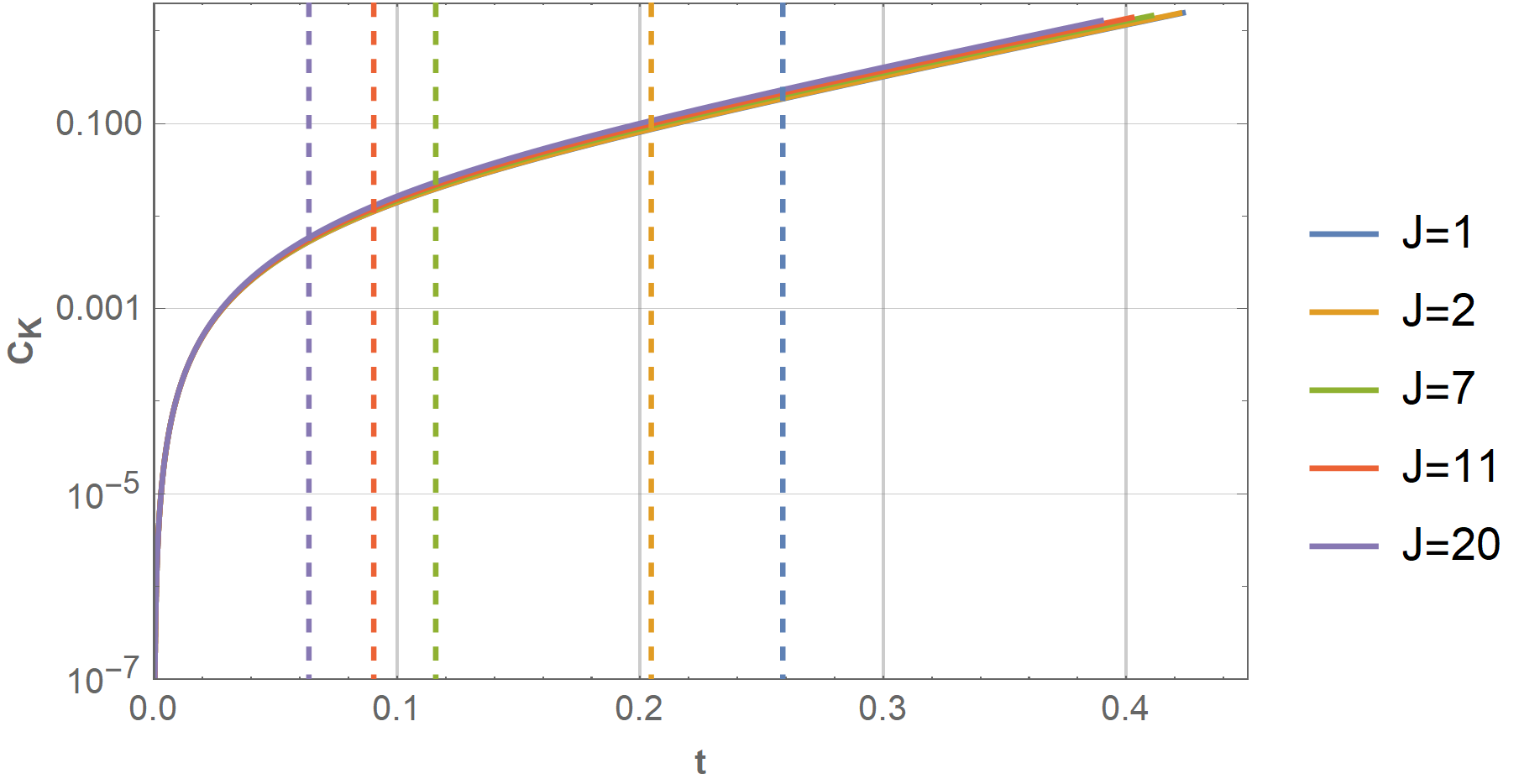}}\\[3pt]
\textnormal{(b) Behavior at late times.}
\end{minipage}
\caption{Krylov complexity $C_K$ versus time at fixed $T=2$ and $z=2.5$ for $J=1,2,7,11,20$. The dashed vertical lines represents the time which the probe traverses the ergosurface.}
\label{fig:KrylovparadistintosJ}
\end{figure}

Figure \ref{fig:KrylovparadistintosJ} shows the spread Krylov complexity as a function of time, for different values of the angular momentum and fixed temperature, and dynamical exponent $z=5/2$. The plots focus at early and late times, and while the former has a linear scale on both axis, the latter has a logarithmic scale on the vertical one. This facilitates the interpretation. At early times, the plots for the complexity are correctly described by a quadratic growth in time, while for late times one observes an exponential growth. We have included the times at which the massive probe enters into the ergoregion, which has no effect on the continuity of the growth of the spread complexity, neither on the continuity of its derivatives. A similar behavior is found when the angular momentum is fixed and the temperature is varied, as depicted in figure \ref{fig:KrylovparadistintosJ}.

\begin{figure}[htbp]
\centering
\begin{minipage}[t]{0.49\linewidth}
\centering
\resizebox{\linewidth}{!}{\includegraphics{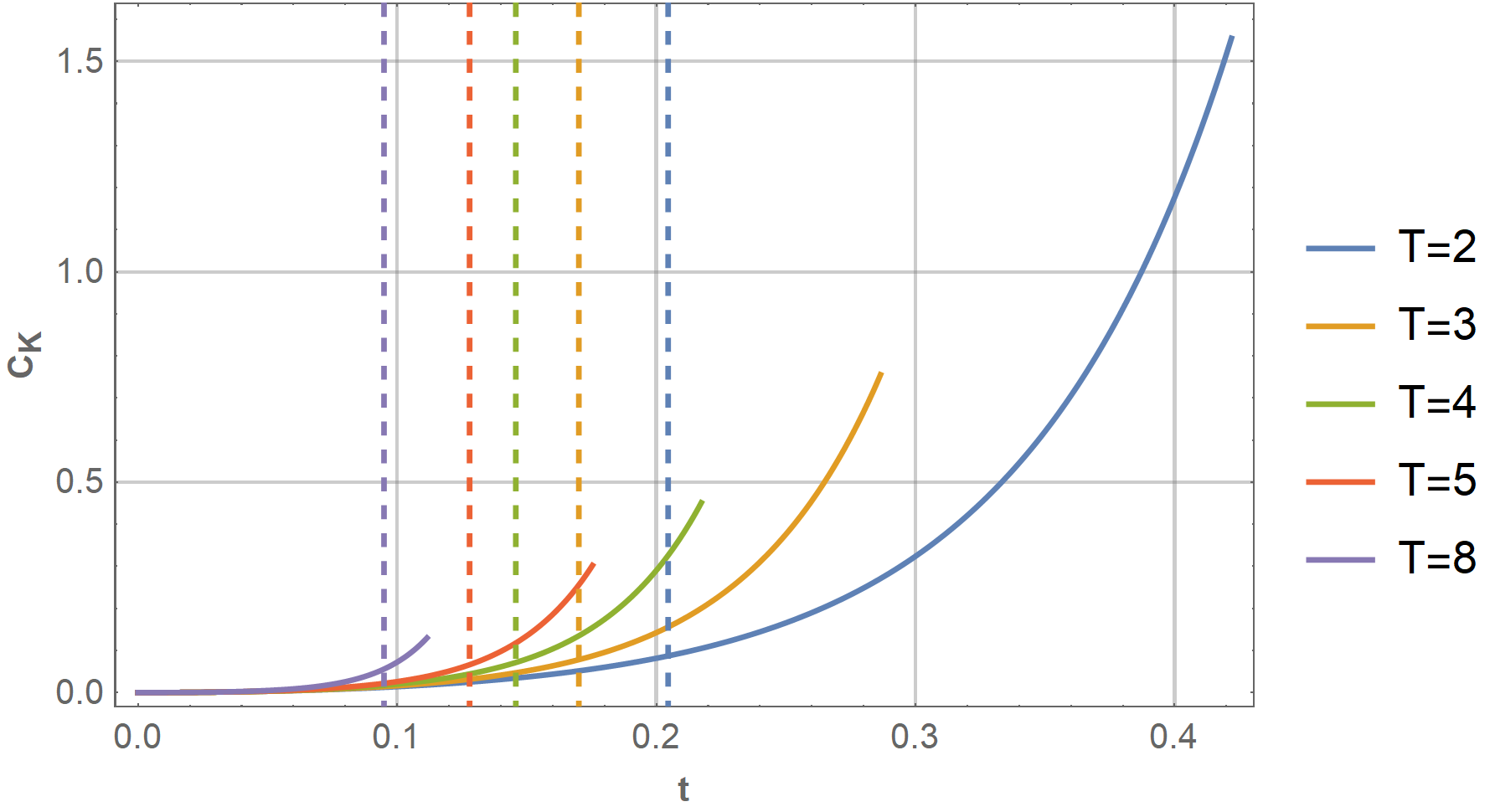}}\\[3pt]
\textnormal{(a) Behavior at early times.}
\end{minipage}\hfill
\begin{minipage}[t]{0.49\linewidth}
\centering
\resizebox{\linewidth}{!}{\includegraphics{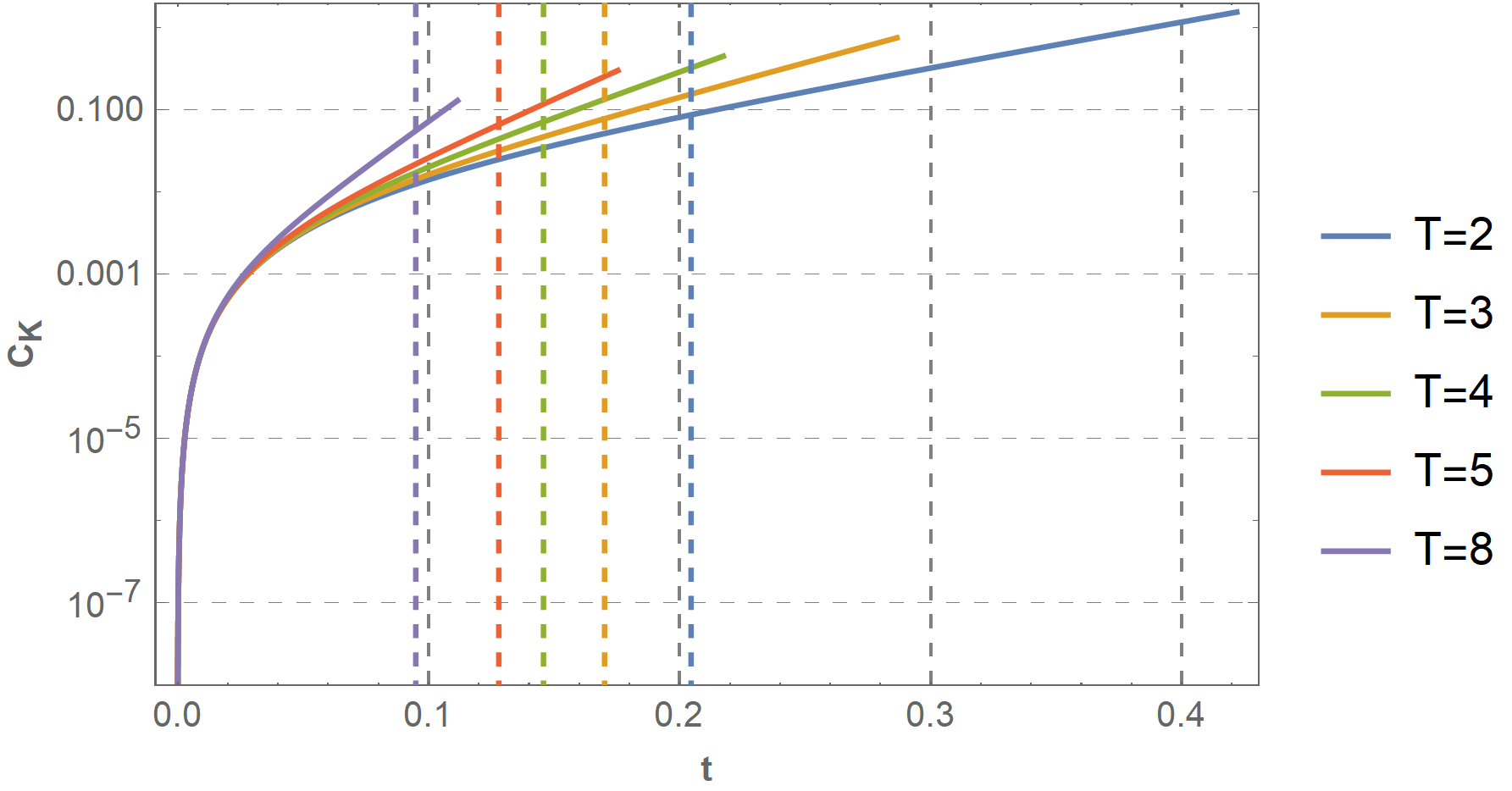}}\\[3pt]
\textnormal{(b) Behavior at late times.}
\end{minipage}
\caption{Krylov complexity $C_K$ versus time at fixed $J=2$ and $z=2.5$ for $T=2,3,4,5,8$. The dashed vertical lines represents the time which the probe traverses the ergosurface.}
\label{fig:KrylovparadistintosT}
\end{figure}
Notice that the late time rate of exponential growth of the complexity is almost insensitive to the angular momentum (see (b) in Figure \ref{fig:KrylovparadistintosJ}), but it changes with the temperature in a significant manner. It can be seen that the late time, exponential growth as a function of the temperature is correctly fitted by $C_\text{late}(t)\sim e^{2\pi T t}$ (see (b) in Figure \ref{fig:KrylovparadistintosT}).
\begin{figure}[H]
   \centering
   \includegraphics[width=0.5\textwidth]{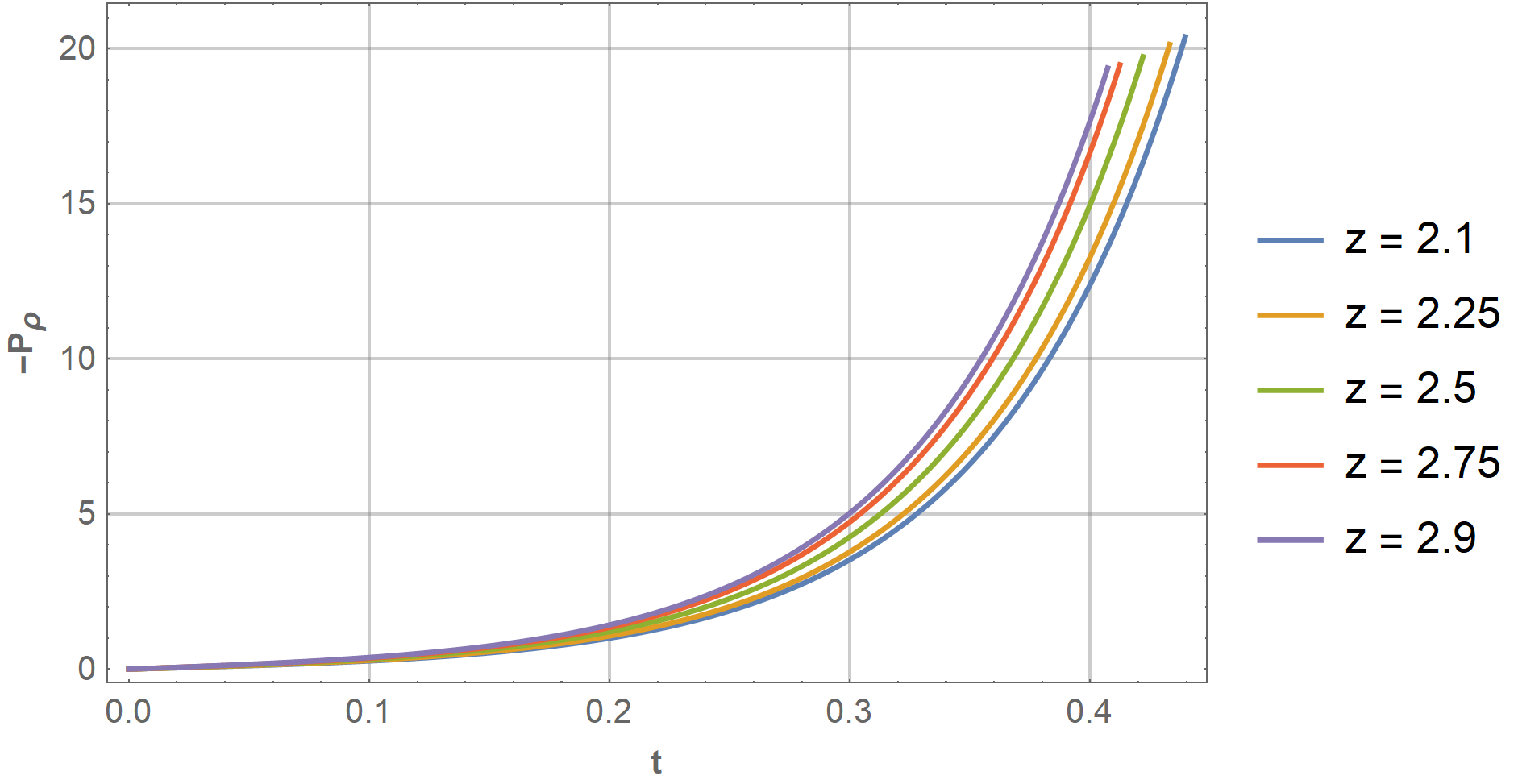}
   \caption{Proper momentum versus time, at $J=2$ and $T=2$, for different values of $z$}
   \label{PvariandoZ}
\end{figure}
Finally, Figure \ref{PvariandoZ} depicts the rate of growth of the spread Krylov complexity as a function of the dynamical exponent $z$, for fixed values of the angular momentum and the temperature.

\section{Conclusions}\label{sec:conc}
We studied two holographic probes on the five-dimensional, asymptotically Lifshitz rotating black hole of \cite{Bravo-Gaete:2026bdq}, with $2<z<3$. Since Lifshitz symmetry has no boosts, the effects of rotation cannot be recovered from a static configuration, and the rotation parameter also controls the electric and axionic charges as extensively explained in \cite{Bravo-Gaete:2026bdq}. The results below therefore describe a correlated change of the state and its supporting fields.

For the static quark-antiquark potential, the pair was separated along the compact rotation direction. The disconnected configuration was taken to be two straight strings ending on a probe brane at the ergosurface. This required the turning point of the connected string to lie outside the ergoregion, since a static worldsheet cannot penetrate it. It is important to notice that the static worlsheet restriction is a consistent one, and it is supported as a static configuration, regardless the dragging, by the string tension. At fixed temperature, the screening length decreases as $J$ increases, which we interpret as a centrifugal effect on the pair. At fixed $J$, it decreases as $T$ increases, as expected from thermal fluctuations weakening correlations at a given separation. At short distances the potential scales as $L^{-z}$, as dictated by anisotropic scaling.

For the spread Krylov complexity, we extended the proposal of \cite{BitaghsirFadafan:2026lek} to the rotating background. Radial motion is not geodesic because of frame dragging, so we eliminated the cyclic coordinate through the Routhian, following \cite{chatzis:2026}. This keeps $P_\rho(0)=0$ and leaves the trajectory unchanged. We found quadratic growth, $C_K\propto t^2$, at early times and exponential growth at late times. The late-time exponent is well fitted by $e^{2\pi T t}$ and is almost insensitive to $J$. Crossing the ergosurface does not affect the continuity of the complexity or its derivatives. At fixed $J$ and $T$, the growth rate at a given instant increases with $z$.

Several directions remain open. Two of them are: 1) a physical explanation of the non-monotonic dependence of $L_0$ on $z$ at fixed $J$, 2) one could consider strings that enter the ergoregion, which requires non-static worldsheets. Finally, other probes, such as entanglement entropy or Wilson loops with different orientations, could be explored in this background.

\section{Acknowledgments}
We thank Carlos Núñez for enlightening comments. J.O. is supported in part by the FONDECYT grant 1262452.
The work of M.C. is partially funded by DGAPA-UNAM grant IG101326. MB is supported by Proyecto Interno UCM-IN-25202
línea regular.

\end{document}